# HyberLoc: Providing Physical Layer Location Privacy in Hybrid Sensor Networks

Rania El-Badry
Wireless Intelligent Networks Center
Nile University, Egypt
Email: rania.elbadry@nileu.edu.eg

Ahmed Sultan
Wireless Intelligent Networks Center
Nile University, Egypt
Email: asultan@nileuniversity.edu.eg

Moustafa Youssef
Wireless Intelligent Networks Center
Nile University, Egypt
Email: mayoussef@nileu.edu.eg

*Abstract*—In many hybrid wireless sensor networks' applications, sensor nodes are deployed in hostile environments where trusted and un-trusted nodes co-exist. In anchor-based hybrid networks, it becomes important to allow trusted nodes to gain full access to the location information transmitted in beacon frames while, at the same time, prevent un-trusted nodes from using this information. The main challenge is that un-trusted nodes can measure the physical signal transmitted from anchor nodes, even if these nodes encrypt their transmission. Using the measured signal strength, un-trusted nodes can still tri-laterate the location of anchor nodes. In this paper, we propose *HyberLoc*, an algorithm that provides anchor physical layer location privacy in anchor-based hybrid sensor networks. The idea is for anchor nodes to dynamically change their transmission power following a certain probability distribution, degrading the localization accuracy at un-trusted nodes while maintaining high localization accuracy at trusted nodes. Given an average power constraint, our analysis shows that the *discretized exponential distribution* is the distribution that maximizes location uncertainty at the un-trusted nodes. Detailed evaluation through analysis, simulation, and implementation shows that *HyberLoc* gives trusted nodes up to 3.5 times better localization accuracy as compared to un-trusted nodes.

## I. INTRODUCTION

Location discovery in wireless sensor networks (WSN) has become a vital field of research due to its critical need in many applications including location-based routing [1], coverage [2], node identification, and information tagging. Localization algorithms can be categorized as either anchor-based or anchor-free [3]. Anchor-based algorithms depend on the presence of a small set of nodes with known locations, i.e. anchor nodes, that broadcast their location information to the network in special *beacon* frames. A node with an unknown location estimates its distance to the anchor node, in a process known as ranging, and combines the estimated distance to at least three anchor nodes to estimate its location in 2D (Fig. 1). On the other hand, anchor-free localization algorithms do not assume the existence of anchor nodes and estimate the relative topology of the network. This paper focuses on anchor-based localization algorithms.

In many applications of wireless sensor networks, sensor nodes are deployed in hostile environments where trusted and un-trusted nodes co-exist. In such hybrid networks, it becomes important to allow trusted nodes to share information while, at the same time, prevent un-trusted nodes from gaining access to this information. In the context of location determination, un-trusted nodes may try to access unauthorized information either to gain access to network resources or to estimate the location of key entities in the network, e.g. anchor nodes. For example, in anchor-based networks, any attack that disables the small set of anchor nodes may render the entire network unoperational.

An anchor node may encrypt its beacon frames with a key shared only with trusted nodes. This will prevent un-trusted nodes from getting the information contained in the beacon frames. Although encryption can provide location information **secrecy**, it does not provide *physical layer location privacy* as an un-trusted node may measure the received signal strength (RSS) of encrypted frames and estimate the distance to anchor nodes with a reasonable accuracy. In addition, un-trusted nodes can collaborate together to determine the location of the anchor node (Fig. 1) based **only** on the measured RSS.

In this paper, we propose the *HyberLoc* algorithm that addresses the physical layer location privacy problem. *HyberLoc* depends on making anchor nodes dynamically and randomly change their transmission power, hence increasing the localization error at un-trusted nodes. At the same time, anchor nodes send the used transmit power encrypted in beacon frames, allowing trusted nodes, that share common information with anchor nodes, to remove the ambiguity of the transmitted power. The shared information can be the key used by anchors to encrypt their beacon frames. The shared information can also be the type, parameters and seed of the probability distribution used by anchors to generate the random transmission power. One main question that our work answers is what the optimal probability distribution that can be used to minimize the localization accuracy at un-trusted nodes under a certain **average power constraint** is.

Previous work in the area of secure localization has focused mainly on two problems: location verification and robust self-location estimation. The goal of location verification techniques, e.g. [4], [5], is to prevent malicious nodes from claiming to be at other locations to gain access to the network resources. On the other hand, the goal of robust self-location estimation techniques, e.g. [6], [7], is for a sensor node to estimate its own location in the presence of attacks, such as malicious anchor nodes. Recently, we addressed another aspect of the the physical layer location privacy problem, which is the un-observability of anchor nodes [8].

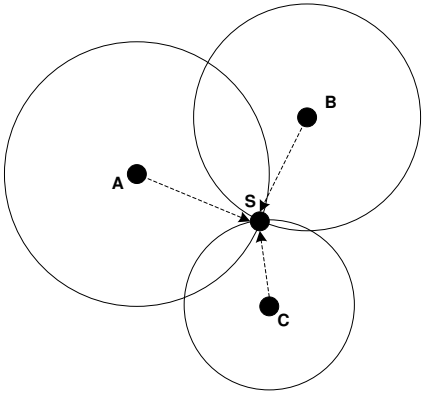

Fig. 1. Node $S$ can estimate its location in 2D using the location beacons received from the three anchor nodes $A$, $B$, and $C$ and the estimated range to them. Similarly, three un-trusted nodes $A$, $B$, and $C$ can cooperate to estimate the location of an anchor node $S$.

The rest of this paper is organized as follows. Section II provides background information. Section III formulates the problem. The *HyberLoc* algorithm is proposed in Section IV. Section V finds optimal probability distribution that can be used to minimize the localization accuracy at un-trusted nodes. Section VI evaluates the performance of the *HyberLoc* algorithm through simulations and implementation. Section VII concludes the paper.

## II. BACKGROUND

### A. Anchor-based vs. Anchor-free Localization

Location discovery algorithms for sensor networks can be classified as either anchor-based or anchor-free. Anchor-based algorithms, e.g. [9], [10], assume that a small percentage of the nodes, i.e. anchor nodes, are aware of their positions. Anchor nodes broadcast their location information to their neighbors which use this information to estimate their own location. In anchor-free algorithms, e.g. [11], [12], no special anchor nodes exist in the network. In this case, the algorithm estimates relative positions, and the coordinate system is established by a reference group of nodes. Relative positioning is suitable for some applications, e.g. location-aided routing [13]. The accuracy of anchor-free algorithms is typically less than anchor-based algorithms. This paper focuses on anchor-based algorithms.

### B. Range Estimation

Ranging is the process of estimating node-to-node distances. In order to determine its location in 2D, a sensor node needs to estimate its range to three or more anchor nodes. Localization methods are mostly based on either Time-of-Arrival (ToA), Angle-of-Arrival methods (AoA), or Received-Signal-Strength (RSS) methods. This paper focuses on the RSS-based range estimation methods where the propagation loss can be calculated based on the difference in power between the transmitted and received signals. Theoretical and empirical models are used to translate this loss into a distance estimate. Combining the positioning information received from at least three anchor nodes with the estimated distances, a sensor node can estimate its location in 2D (Fig. 1).

### C. Cramer-Rao Lower Bound

In the field of estimation, the Cramer-Rao bound is a lower bound on the covariance of an unbiased estimator. Any estimator achieves this lower bound is said to be an efficient estimator. In some cases, no estimator exists which achieves this lower bound.

Cramer-Rao bound applies only to unbiased estimators [14]. In other words, there may exist a biased estimator that achieves a lower variance than Cramer-Rao lower bound. The Cramer-Rao lower bound (CRLB) is the inverse of the Fisher information matrix $I(\theta)$. The element $(m, n)$ in $I(\theta)$ is given by:

$$I_{m,n}(\theta) = -E[\frac{\partial^2}{\partial \theta_m \partial \theta_n} \log f(Z|\theta)] \quad (1)$$

where $\theta$ is the vector of parameters, $\theta_k$ is the $k^{th}$ parameter, $Z$ is the vector of observations, and $f(Z|\theta)$ is the likelihood function.

## III. PROBLEM FORMULATION

This section outlines the system model, security requirements, and signal model.

### A. System Model

We assume a hybrid wireless sensor network where anchors, trusted and un-trusted nodes co-exist. Anchor nodes periodically broadcast their encrypted beacon frames. The power levels used by anchors are chosen following a certain probability distribution with a constraint on the average transmission power. All nodes, including un-trusted nodes, are assumed to use the same hardware and use received signal strength for distance estimation.

Furthermore, we assume that un-trusted nodes are passive, i.e. they do not inject any traffic, and have full access to the network traffic. In addition, we assume that anchor nodes and trusted nodes are not compromised. The goal of the un-trusted nodes is to estimate the distance to anchor nodes based on the physical signal transmitted by these nodes. Finally, we assume that un-trusted nodes use the Maximum Likelihood (ML) estimation method to obtain transmission power levels and estimate their locations.

### B. Security Requirements

We have two main security requirements that should be satisfied.

*1) Location Information Secrecy:* Anchor nodes should be able to broadcast their position information periodically and trusted nodes should be able to use this information to estimate their position. On the other hand, un-trusted nodes should not be able to use anchor nodes' beacon frames to gain information about anchor nodes' locations. This can be achieved, for example, by encrypting the anchor nodes' beacons.

*2) Physical Layer Location Privacy:* Un-trusted nodes should not be able to exploit the measured physical signal (RSS) to estimate the location of anchor nodes. This paper focuses on this security requirement.

## C. Signal Model

We consider a signal propagation model that has a dominant line-of-sight (LOS) component. In the presence of Additive White Gaussian Noise (AWGN), the probability density function (PDF) of the received signal power [15] is:

$$f(z|h,x) = \frac{1}{2\sigma^2}\exp(-\frac{z+hx}{2\sigma^2})I_0(\frac{\sqrt{zhx}}{\sigma^2}) \quad (2)$$

where
- $z$: is the received signal power.
- $h$: is the channel gain which is a function of distance.
- $x$: is the transmission power.
- $2\sigma^2$: is the total variance of noise.
- $I_0(x)$: is the modified Bessel function of order 0.

For a sensor node employing the ML estimation method to determine the distance to a certain transmitting node, the likelihood function ($K$) using $m$ independent measurements is:

$$K = \prod_{k=1}^{m} \frac{1}{2\sigma^2}\exp(-\frac{z_k+hx_k}{2\sigma^2})I_0(\frac{\sqrt{z_k h x_k}}{\sigma^2}) \quad (3)$$

The ML technique is based on finding $h$ and $x_k$ that maximize $K$.

## IV. HYBERLOC ALGORITHM

This section provides a detailed description of the proposed *HyberLoc* algorithm and the optimal probability distribution that could be used by the anchors.

### A. Algorithm

The proposed *HyberLoc* algorithm addresses the physical layer location privacy mentioned in section III-B. Before transmitting a beacon frame, the anchor node chooses a random transmit power $x$, includes it in the beacon frame, and transmits the encrypted beacon frame using the selected transmit power $x$.

When an un-trusted node receives a beacon frame, it cannot determine the transmit power $x$ and hence has higher ambiguity which leads to lower accuracy in location estimation. We want to emphasize here that encryption alone, without changing the transmit power, is not enough as three un-trusted nodes can cooperate to determine the location of the anchor node, based on the measured RSS at the physical layer, as shown in Fig. 1.

On the other hand, a trusted node receiving a beacon frame can use the shared encryption key with the anchor node to get the transmit power, $x$, that was included in the frame, removing the ambiguity introduced by changing the transmit power.

In the next section, we analytically find the probability distribution that could be used by anchors to minimize the localization accuracy at un-trusted nodes.

## V. OPTIMAL DISTRIBUTION

### A. Statistical Approach

One way of finding the probability distribution that minimizes the localization accuracy at un-trusted nodes is to find the probability distribution that maximizes the localization variance at un-trusted node. Thus, we can find the CRLB for the estimated distance at un-trusted nodes for every probabilistic distribution. Then, choose the distribution with the highest lower bound.

*1) Analysis:* As a result of the direct relationship between the estimated channel gain ($h$) and the estimated distance, the variance of the estimated channel gain perfectly reflects the variance of the estimated distance.

To find the CRLB for the un-trusted estimated channel gain, we calculate Fisher information matrix. For un-trusted nodes, the unknowns are the channel gain $h$ and the transmission power sequence $x_k \ \forall \ k = 1, 2, ......, m$. In this case, the un-trusted should estimate $m+1$ parameters. Thus, the Fisher information matrix is a $m+1 \times m+1$ matrix.

For the Likelihood function in equation 3, $I_{i,j}(\theta)$ can be calculated using:

1) For $i = 1$ and $j = 2, 3, ....., m+1$

$$\frac{\partial^2}{\partial h \partial x_{j-1}} \log K = \frac{-1}{4}[\frac{2\sigma^2 I_0(\frac{\sqrt{z_{j-1}hx_{j-1}}}{\sigma^2})^2}{\sigma^4 I_0(\frac{\sqrt{z_{j-1}hx_{j-1}}}{\sigma^2})^2} \quad (4)$$

$$-\frac{z_{j-1}I_0(\frac{\sqrt{z_{j-1}hx_{j-1}}}{\sigma^2})^2}{\sigma^4 I_0(\frac{\sqrt{z_{j-1}hx_{j-1}}}{\sigma^2})^2}$$

$$+\frac{z_{j-1}I_1(\frac{\sqrt{z_{j-1}hx_{j-1}}}{\sigma^2})^2}{\sigma^4 I_0(\frac{\sqrt{z_{j-1}hx_{j-1}}}{\sigma^2})^2}]$$

2) For $i = 1$ and $j = 1$

$$\frac{\partial^2}{\partial h^2} \log K = \sum_{n=1}^{m} \frac{-1}{4}z_n x_n[\frac{-\sqrt{(z_n h x_n)}I_0(\frac{\sqrt{z_n h x_n}}{\sigma^2})^2}{h\sqrt{z_n h x_n}\sigma^4 I_0(\frac{\sqrt{z_n h x_n}}{\sigma^2})^2} \quad (5)$$

$$+\frac{2\sigma^2 I_0(\frac{\sqrt{z_n h x_n}}{\sigma^2})I_1(\frac{\sqrt{z_n h x_n}}{\sigma^2})}{h\sqrt{z_n h x_n}\sigma^4 I_0(\frac{\sqrt{z_n h x_n}}{\sigma^2})^2}$$

$$+\frac{I_1(\frac{\sqrt{z_n h x_n}}{\sigma^2})^2 \sqrt{z_n h x_n}}{h\sqrt{z_n h x_n}\sigma^4 I_0(\frac{\sqrt{z_n h x_n}}{\sigma^2})^2}]$$

3) For $i = 2, ..., m+1$ and $j = 1$

$$\frac{\partial^2}{\partial x_{i-1}\partial h}\log K = \frac{-1}{4}\Big[\frac{2\sigma^2 I_0(\frac{\sqrt{z_{i-1}hx_{i-1}}}{\sigma^2})^2}{\sigma^4 I_0(\frac{\sqrt{z_{i-1}hx_{i-1}}}{\sigma^2})^2} \quad (6)$$

$$-\frac{z_{i-1}I_0(\frac{\sqrt{z_{i-1}hx_{i-1}}}{\sigma^2})^2}{\sigma^4 I_0(\frac{\sqrt{z_{i-1}hx_{i-1}}}{\sigma^2})^2}$$

$$+\frac{z_{i-1}I_1(\frac{\sqrt{z_{i-1}hx_{i-1}}}{\sigma^2})^2}{\sigma^4 I_0(\frac{\sqrt{z_{i-1}hx_{i-1}}}{\sigma^2})^2}\Big]$$

4) For $i = j = 2, ..., m+1$

$$\frac{\partial^2}{\partial x_{j-1}{}^2}\log K = \quad (7)$$

$$\frac{-hz_{j-1}}{4}\Big[\frac{-\sqrt{(z_{j-1}hx_{j-1})}I_0(\frac{\sqrt{z_{j-1}hx_{j-1}}}{\sigma^2})^2}{x_{j-1}\sqrt{z_{j-1}hx_{j-1}}\sigma^4 I_0(\frac{\sqrt{z_{j-1}hx_{j-1}}}{\sigma^2})^2}$$

$$+\frac{2\sigma^2 I_0(\frac{\sqrt{z_{j-1}hx_{j-1}}}{\sigma^2})I_1(\frac{\sqrt{z_{j-1}hx_{j-1}}}{\sigma^2})}{x_{j-1}\sqrt{z_{j-1}hx_{j-1}}\sigma^4 I_0(\frac{\sqrt{z_{j-1}hx_{j-1}}}{\sigma^2})^2}$$

$$+\frac{I_1(\frac{\sqrt{z_{j-1}hx_{j-1}}}{\sigma^2})^2\sqrt{z_{j-1}hx_{j-1}}}{x_{j-1}\sqrt{z_{j-1}hx_{j-1}}\sigma^4 I_0(\frac{\sqrt{z_{j-1}hx_{j-1}}}{\sigma^2})^2}\Big]$$

5) For $i \neq 1$, $j \neq 1$ and $i \neq j$

$$\frac{\partial^2}{\partial x_i \partial x_j}\log K = 0 \quad (8)$$

*2) Results:* Using the above information matrix and averaging over sufficient number of iterations, Fig. 2 shows the CRLB for trusted and un-trusted estimated channel gain. Note that the for trusted nodes, the CRLB is much lower than un-trusted nodes' CRLB as trusted nodes know exactly the transmission power sequence used by anchors so they will only have to estimate the channel gain $h$. For un-trusted nodes, they will have to estimate not only the channel gain but also the the transmission power sequence $x_k \ \forall \ k = 1, ..., m$.

*3) Discussion:* Fig. 2 shows that CRLB approach does not guide us to the optimal probabilistic distribution that should be used by anchors to maximize the estimated distance variance measured at the un-trusted nodes. As is evident in the Fig. 2, CRLB is almost the same regardless of distribution. However, this approach shows the promise of the proposed *HyberLoc* algorithm in degrading the localization accuracy at un-trusted nodes.

In the next section, we use an information theoretic approach to find the variance-maximizing probabilistic distribution.

*B. Information Theoretic Approach*

As a criterion for selecting a probability distribution of transmission power, we choose to find the distribution that maximizes the entropy (uncertainty) [16] over all possible power levels subject to an average power constraint ($\mu$). Based on the current sensor network hardware, e.g. [17], transmission power can be selected from a set of pre-specified discrete power levels. We formulate the problem as a convex optimization problem that could be solved using Lagrange multipliers.

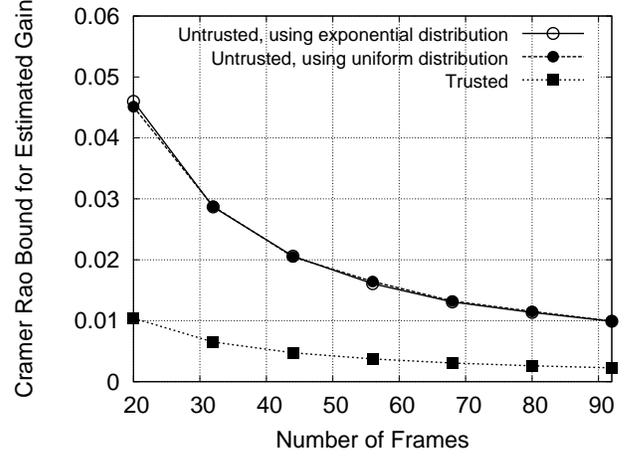

Fig. 2. The Cramer-Rao Lower Bound for trusted and un-trusted nodes using exponential and normal distribution

*1) Notations:*
- $x$: is a discrete random variable representing the power level used by anchors.
- $n$: is the number of possible power levels.
- $\mu$: is a target average power that anchors are designed to use.
- $P(x)$: is the probability distribution used by anchors to generate $x$.
- $H(x)$: is the entropy of the probability distribution $P(x)$ given by:

$$H(x) = -\sum_x P(x)\log P(x) \quad (9)$$

*2) Optimization Problem:*

$$\max. -\sum_{i=1}^n P(x_i)\log P(x_i) \quad (10)$$

$$s.t.: \sum_{i=1}^n P(x_i) = 1 \quad (11)$$

$$\sum_{i=1}^n x_i P(x_i) = \mu \quad (12)$$

We define the Lagrangian, $L$, as:

$$L = -\sum_{i=1}^n P(x_i)\log P(x_i) + \lambda_1 \sum_{i=1}^n P(x_i) + \lambda_2 \sum_{i=1}^n x_i P(x_i) \quad (13)$$

where $\lambda_1$ and $\lambda_2$ are Lagrange multipliers. This problem is convex and the optimal $P(x_i)$ can be obtained by differenti-ating $L$ with respect to $P(x_i)$ and equating to zero.

$$\frac{\partial L}{\partial P(x_i)} = -\log P(x_i) - 1 + \lambda_1 + \lambda_2 x_i = 0 \quad (14)$$

$$P(x_i) = \exp(\lambda_1 - 1)\exp(\lambda_2 x_i) \quad (15)$$

$$P(x_i) = k\exp(-\alpha x_i) \quad (16)$$

From constraints 11 and 12:

$$\sum_{i=1}^{n} k\exp(-\alpha x_i) - 1 = 0 \quad (17)$$

$$\sum_{i=1}^{n} x_i k\exp(-\alpha x_i) - \mu = 0 \quad (18)$$

By solving Equations 17 and 18 simultaneously, the values for $k$ and $\alpha$ are determined.

*3) Discussion:* Equation 16 shows that the discretized exponential distribution is the entropy-maximizing distribution when there is a constraint on the average transmission power. Thus, when the anchor nodes use a discretized exponential distribution, they make it harder for un-trusted nodes to find the $x_k$ sequence used, hence, minimize their localization accuracy. On the other hand, the localization accuracy at trusted nodes is not affected by the type of the distribution as the transmit power is encrypted in the frame. Note that if the average power is not constrained, the uniform distribution would be the entropy-maximizing distribution.

## VI. PERFORMANCE EVALUATION

In this section, we evaluate the performance of the proposed *HyberLoc* algorithm using simulation and implementation. Our goal is to show that when a proper probabilistic distribution is used by anchors to randomize their transmission power, a degradation in localization accuracy at un-trusted nodes is achieved while the localization accuracy at trusted nodes is not affected. We evaluate the performance under three different probabilistic distributions: uniform, discretized normal, and discretized exponential distributions (optimal distribution).

### A. Performance Metric

The performance metric used is the standard deviation of estimated distance error normalized by the true distance. It reflects the localization accuracy over a sufficient number of samples (frames). Our goal is to keep this metric as high as possible for un-trusted nodes. On the other hand, we should guarantee a reasonable anchor localization accuracy for trusted nodes.

### B. Simulation Experiment

*1) Environment:* The algorithm is implemented using Matlab. Trusted and un-trusted nodes are placed at equal distance from the anchor node. Since our goal is to quantify the performance advantage of trusted nodes over un-trusted nodes, without loss of generality, all experiments have been performed using one anchor node, one trusted node and one un-trusted node only. Anchors are assumed to have four possible power levels[1]. As previously mentioned, trusted and un-trusted nodes use ML to localize the anchor node but only trusted nodes know exactly the power level used for every frame. That is, the unknown parameters to be estimated in Equation 3 are $h$ for trusted nodes, and $h$ and all $x_k$ for un-trusted nodes. For un-trusted nodes, the optimal $x_k$ sequence is obtained via exhaustive search. In order to reduce the domain of exhaustive search, the un-trusted nodes divide all received frames into blocks of size $s$, exhaustively search for the power level sequence that maximizes $K$, and then average the estimates for $h$ obtained from each block. In our simulations, $s$ is chosen to be four. Larger $s$ does not increase the accuracy significantly but increases the search time exponentially.

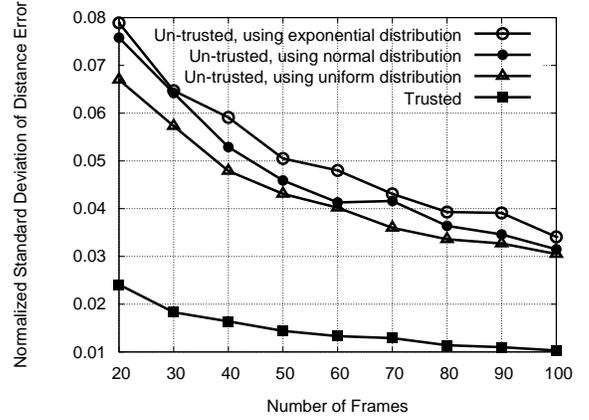

(a) Average SNR = 16 dB.

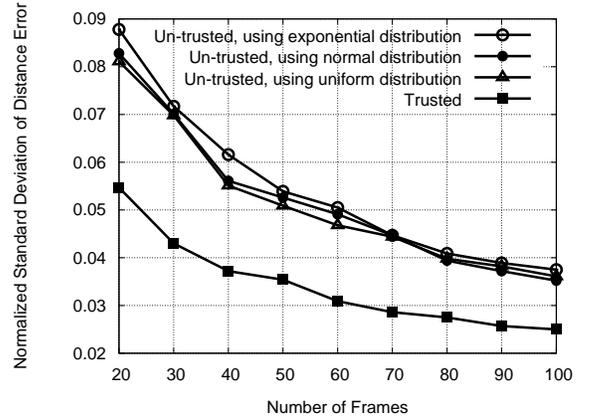

(b) Average SNR = 9 dB.

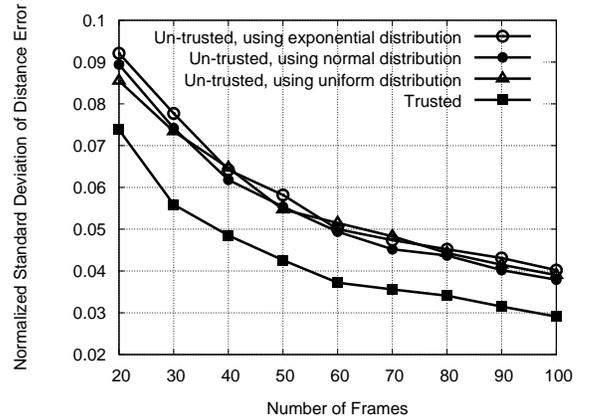

(c) Average SNR = 6.5 dB.

Fig. 3. Normalized standard deviation of distance error as a function of the number of frames used for estimating the distance using different probability distributions.

---

[1]Note that typical sensor nodes, e.g. [17], can have up to 31 power levels. Since adding more power levels will increase the search time significantly at un-trusted nodes, our results here present a lower bound on the enhancement obtained using the *HyberLoc* algorithm.

|           | Min. | Avg. | Max. |
|-----------|------|------|------|
| SNR= 16 dB | 3.2  | 3.4  | 3.5  |
| SNR= 9 dB  | 1.4  | 1.5  | 1.6  |
| SNR= 6.5 dB| 1.2  | 1.3  | 1.4  |

TABLE I
RATIO OF LOCALIZATION ACCURACY ENHANCEMENT BETWEEN THE TRUSTED NODE AND UN-TRUSTED NODE WHEN USING THE *HyberLoc* ALGORITHM FOR DIFFERENT SNR.

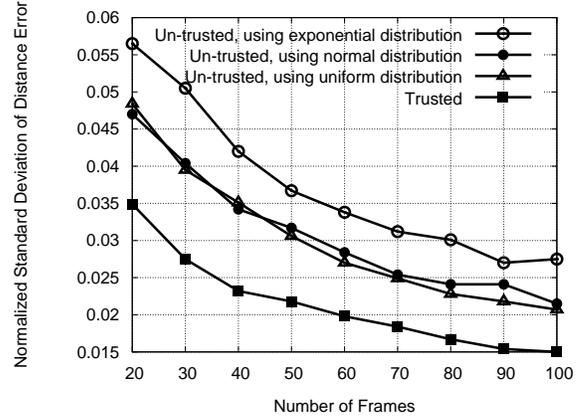

Fig. 4. Results of normalized standard deviation of estimated distance error via implementing HyberLoc on TelosB motes.

We evaluate the performance of the *HyberLoc* algorithm under different signal to noise ratio (SNR) levels. Note that since our metric is normalized by the actual distance between a node and the anchor node, the absolute distance is not included as a parameter.

*2) Results:* Fig. 3 provides the normalized standard deviation of estimated distance error for different probability distributions of transmission power, obtained via Monte Carlo simulations, as a function of the number of frames used for estimating the distance for different levels of SNR. The average power, $\mu$, is set to -3 dBm. The figure verifies that the discretized exponential distribution maximizes the normalized standard deviation of estimated distance error. The figure also shows that using the *HyberLoc* algorithm, the localization accuracy at trusted nodes is at least 3.2, 1.4, and 1.2 times better than that at un-trusted nodes for SNR= 16, 9, and 6.5 dB respectively. As the number of frames used in estimation increases, the localization accuracy at both trusted and un-trusted nodes increases.

As the SNR increases, the difference between the different distributions increases as well as the difference between trusted and un-trusted nodes.

Table I summarizes the results.

### C. Implementation

*1) Environment:* We have implemented our algorithm on TelosB motes [17] in an indoor environment replicating the configuration used in Matlab simulations where the anchor-trusted distance and anchor-untrusted distance are 1m. The anchor node periodically broadcast its beacon frames. Both trusted and un-trusted nodes measure the received signal strength (RSS) and use it for distance estimation.

*2) Results:* Fig. 4 shows the effect of using the *HyberLoc* algorithm on the normalized standard deviation of estimated distance error using different probability distributions. The figure confirms that when the anchor uses a discretized exponential distribution, the maximum degradation in the localization accuracy at the un-trusted node is achieved.

Unlike the simulation results (Fig. 3), the figure shows some crossovers between uniform and normal distributions curves as a result of the multi path fading encountered in indoor environments.

### D. Summary

In this section, we evaluated the performance of the proposed *HyberLoc* algorithm through simulations and implementation. The results show that the *HyberLoc* algorithm can cause degradation in the localization accuracy at un-trusted nodes without affecting the accuracy at trusted nodes. The results also show that the discretized exponential distribution minimizes the accuracy at un-trusted nodes validating the analytical results in Section V.

## VII. CONCLUSION

In this paper, we focused on the physical layer location privacy problem, where an anchor node is required to hide its physical location from un-trusted nodes. We have proposed the *HyberLoc* algorithm for solving the physical layer location privacy and evaluated its performance through analysis, simulations and implementation. We proved analytically that the discrete exponential distribution is the entropy-maximizing distribution under a given average transmit power constraint. Our results show that the *HyberLoc* algorithm can cause degradation in localization accuracy at un-trusted nodes (up to 3.5 times worse) without limiting the localization accuracy at trusted nodes. In addition, it has a low overhead and does not need any additional hardware, making it suitable for the resource-constrained sensor networks.